\documentclass[runningheads]{llncs}

\usepackage[utf8]{inputenc}
\usepackage{graphicx}
\usepackage[colorlinks=true,bookmarksopen,bookmarksdepth=2]{hyperref}

\usepackage{amsmath, amsfonts, amssymb, stmaryrd}
\usepackage{tikz}
\usetikzlibrary{arrows,automata,positioning}

\newcommand{\indsub}[3]{ {#3[ #1 := #2 ]} }
\newcommand{\indsft}[3]{ {\uparrow^{#1}_{\geq #2} #3 } }

\newcommand \alt{\;\;|\;\;}
\newcommand \twnf{\textit{wnf}}
\newcommand \tnf{\textit{nf}}
\newcommand \nf{T_\tnf}
\newcommand \neu{T_{\textit{neu}}}

\newcommand \wnf{T_{W}}
\newcommand \inert{T_{{I}}}
\newcommand \bwnf{\beta\twnf}

\newcommand{\semsub}[2]{\llbracket #1 \rrbracket_{#2}}
\newcommand{\plg}{\operatorname{plug}}
\newcommand{\open}[2]{\operatorname{open}_{#1}(#2)}

\newcommand{\openT}[1]{\open T {#1}}
\newcommand{\openW}[1]{\open W {#1}}
\newcommand{\openI}[1]{\open I {#1}}
\newcommand{\openE}[1]{\open E {#1}}

\newcommand\setofwnf{\mathbf{Wnfs}}
\newcommand\setofinert{\mathbf{Inerts}}
\newcommand{\nil}{\bullet}
\newcommand{\cons}[2]{#1 :: #2}

\newcommand{\tvar}[1]{{#1}}
\newcommand{\tlam}[2]{\lambda{#1}.{#2}}

\newcommand{\tolam}[1]{{\lambda{#1}}}

\newcommand{\tapp}[2]{{#1}\;{#2}}
\newcommand{\tappp}[2]{\tapp{#1}{#2}}
\newcommand{\lvar}[1]{\ensuremath{V}(#1)}

\newcommand{\clos}[2]{{[ #1, #2 ]}}
\newcommand{\abstr}[2]{[\lambda #1, #2]}

\newcommand{\strongWannotation}[1]{\{#1\}_\mathsf{w}}
\newcommand{\strongNannotation}[1]{\{#1\}_\mathsf{n}}
\newcommand{\rlex}{<_\mathit{rlex}}
\newcommand{\lex}{<_\mathit{lex}}

\newcommand{\kwf}{K_{\mathit{wf}}}

\newcommand{\flclos}[2]{\tappp{\clos{#1}{#2}}{\hole}}
\newcommand{\frclos}[2]{\tappp{\hole}{\clos{#1}{#2}}}
\newcommand{\flam}{{\lambda\hole}}
\newcommand{\flstrongann}[1]{\tappp{\strongWannotation{#1}}{\hole}}
\newcommand{\frstrongann}[1]{\tappp{\hole}{\strongNannotation{#1}}}
\newcommand{\flstrong}[1]{\tappp{{#1}}{\hole}}
\newcommand{\frstrong}[1]{\tappp{\hole}{{#1}}}
\newcommand{\flapp}[1]{\tappp{#1}{\hole}}
\newcommand{\frapp}[1]{\tappp{\hole}{#1}}

\newcommand{\ctxnil}{\hspace{0.4mm}\nil\hspace{0.4mm}} 

\newcommand{\econf}[4]{{\langle#1, #2, #4, #3\rangle}}
\newcommand{\wconf}[4]{{\langle#1, #2, #3, #4\rangle_{\cal E}}}
\newcommand{\bconf}[2]{{\langle#2, #1, m\rangle_{\cal C}}}
\newcommand{\bconfone}[2]{{\langle#2, #1, 1\rangle_{\cal C}}}
\newcommand{\bconfwiththreearguments}[3]{{\langle#2, #1, #3\rangle_{\cal C}}}
\newcommand{\knsconf}[3]{{\langle#3, #1, #2\rangle}}  
\newcommand{\sconf}[3]{{\langle#3,  {#1}, #2\rangle_{\cal{S}}}}

\newcommand{\hole}{\Box}

\newcommand{\stackone}[1]{[#1]}
\newcommand{\stacktwo}[2]{[#1;\; #2]}
\newcommand{\stackthree}[3]{[#1;\; #2;\; #3]}
\newcommand{\stackconstwo}[2]{#1::#2::\nil}

\newcommand{\ie}{i.e.}
\newcommand{\eg}{e.g.}

\newcommand{\rightcounter}[1]{\setcounterref{equation}{#1}%
  \addtocounter{equation}{-1}}

\usepackage{refcount}
\usepackage{listings}

\begin{document}
\title{An Abstract Machine for Strong Call by Value\thanks
{This research is supported by the National Science
    Centre of Poland, under grant number 2019/33/B/ST6/00289.}}

\renewcommand{\orcidID}[1]{\href{https://orcid.org/#1}{\includegraphics[width=8pt]{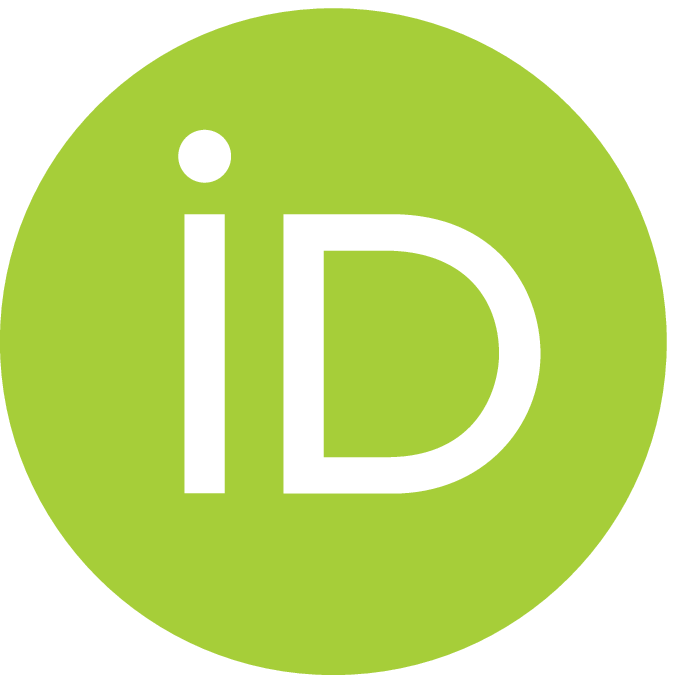}}}

\author{Małgorzata Biernacka\inst{1}\orcidID{0000-0001-8094-0980} \and
Dariusz Biernacki\inst{1}\orcidID{0000-0002-1477-4635} \and
Witold Charatonik\inst{1}\orcidID{0000-0001-7062-0385} \and
Tomasz Drab\inst{1}\orcidID{0000-0002-6629-5839}}
\authorrunning{M. Biernacka et al.}

\institute{Institute of Computer Science, University of Wroc{\l}aw, Poland\\
\email{\{mabi,dabi,wch,tdr\}@cs.uni.wroc.pl}\\
}

\maketitle

\begin{abstract}
  We present an abstract machine that implements a full-reducing
  (a.k.a. strong) call-by-value strategy for pure $\lambda$-calculus.
  It is derived using Danvy et al.'s functional correspondence from
  Crégut's KN by: (1) deconstructing KN to a call-by-name
  normalization-by-evaluation function akin to Filinski and Rohde's,
  (2) modifying the resulting normalizer so that it implements the
  right-to-left call-by-value function application, and (3)
  constructing the functionally corresponding abstract machine.

  This new machine implements a reduction strategy that subsumes the
  fireball-calculus variant of call by value studied by Accattoli et
  al.  We describe the strong strategy of the machine in terms of a
  reduction semantics and prove the correctness of the machine using a
  method based on Biernacka et al.'s generalized refocusing. As a
  byproduct, we present an example application of the machine to
  checking term convertibility by discriminating on the basis of their
  partially normalized forms.


\end{abstract}


\section{Introduction}
\label{sec:intro}

Full-reducing (also known as strong) normalization strategies in the
lambda calculus have so far received relatively little attention
compared to weak strategies that provide foundations for functional
programming languages, such as OCaml (implementing call by value) or
Haskell (implementing call by need).
However, recent advances in proof
technology and the use of proof assistants based on dependently typed
lambda calculus for complex verification efforts propel the design and
study of strong reduction strategies, and of their corresponding
efficient realizations on a
machine~\cite{DBLP:journals/scp/AccattoliG19,Balabonski-al:ICFP17,Gregoire-Leroy:ICFP02}.

Abstract machines provide a convenient computation model that mediates
between the specific reduction strategy in the calculus and its
practical implementations. The first machine for strong normalization
of lambda terms is due to Crégut\cite{Cregut:HOSC07}. This machine
implements normal-order strategy~\cite{DBLP:journals/jfp/Garcia-PerezN19},
\ie, a hybrid strategy that iterates
call by name (CbN), and necessarily extends reduction to open terms
and reduces under lambda abstractions---unlike machines for weak
strategies that operate on closed terms and stop at lambdas. Similarly
to strong CbN, one can define strong call by value (CbV) as an
iteration of weak CbV, carefully generalizing the notion of value to
open terms~\cite{AccattoliG16}. A normalization function realizing
strong CbV was proposed by Gr\'egoire\&Leroy and implemented in their
virtual machine extending the ZAM
machine~\cite{Gregoire-Leroy:ICFP02}. Another virtual machine for
strong CbV was derived by Ager et al.~\cite{Ager-al:RS-03-14} from
Aehlig and Joachimski's normalization
function~\cite{Aehlig-Joachimski:MSCS04}. Recently, a strong
call-by-need strategy has been proposed by Kesner et
al.~\cite{Balabonski-al:ICFP17}, and the corresponding abstract machine
has been derived by Biernacka et al.~\cite{BiernackaC19}. On the other
hand, there is a line of work done by Accattoli et al. who study
computational complexity of abstract machines, in particular in the context of a
weak CbV strategy that operates on open terms, as an intermediate step
towards an efficient machine for strong
CbV~\cite{DBLP:journals/scp/AccattoliG19}.

Many abstract machines are devised or tailored by hand, and their
correctness is far from obvious. Alternatively, Danvy et al. initiated
a derivational approach that allows to obtain abstract machines from
preexisting semantic artefacts for specific strategies by applying
well-defined transformations in a systematic way.

Danvy et al.'s functional correspondence~\cite{Ager-al:PPDP03} is a
two-way semantics-preserving derivation method that relates
higher-order evaluators and abstract machines. More precisely,
following Reynolds' recipe of applying a CPS translation and
defunctionalization to a higher-order evaluator expressed in a
functional meta-language, it leads to an implementation, in the same
meta-language, of the corresponding abstract
machine~\cite{Reynolds:HOSC98}. However, the two program
transformations can be inverted and, as first observed by
Danvy~\cite{Danvy:IFL04}, starting with an implementation of an abstract
machine, one can obtain a higher-order compositional evaluator, in the
style of a valuation function of denotational
semantics~\cite{Schmidt:86}, that abstractly and concisely embodies
the low-level intricacies of the machine, typically scattered all
over the transition rules. Such an evaluator can then be locally
modified according to one's needs and a new abstract machine can be
derived from it. This approach has proven extremely successful at
numerous occasions and it appears to be considerably more systematic
and effective than groping for the right changes directly at the level
of the abstract machine.

The goal of this work is to derive an abstract machine
that can be seen as a strong CbV counterpart
 of Crégut's machine for normal order
 which avoids needless reevaluation of function arguments. 
 Rather than directly tweaking
 the KN machine we propose to take a systematic approach following
 Danvy's recipe and
 (1) we first deconstruct KN into a compositional evaluator, (2) we
 then modify this evaluator accordingly to account for CbV, and (3)
 from the new evaluator we derive a new abstract machine. Our
 meta-language is a small subset of OCaml~\cite{Leroy-al:OCaml-4.10}.

 In the process, we identify the reduction semantics of the rrCbV
 variant of a strong CbV strategy in the pure lambda calculus which we
 also present. In the terminology of Biernacka et al.'s, this is a
 hybrid strategy that uses three kinds of contexts, and it subsumes as
 a substrategy the weak right-to-left strategy of Accattoli et al.'s
 fireball calculus.  As an application of the machine we also show how
 to check convertibility of terms by their partial
 normalization where we can stop the machine and compare computed
 prefixes of normal forms. Thus, the contributions of this paper
 include:
\begin{enumerate}
\item a full systematic derivation of the machine from a CbV evaluator,
\item a presentation of an abstract machine for strong CbV that is a
  counterpart of Crégut's KN machine,
\item a reduction semantics for a strong CbV strategy,
\item an application of the machine to convertibility checking.
\end{enumerate}
\paragraph{Outline.} In Section~\ref{sec:deconstruction} we recall the
KN machine and present the NbE function obtained by its
deconstruction. In Section~\ref{sec:cbv-machine} we present the
machine derived from the evaluator, and in
Section~\ref{sec:reduction-semantics} the corresponding reduction
semantics. In Section~\ref{sec:correctness} we prove the  correctness of the machine 
with respect to the semantics,
and in Section~\ref{sec:conclusion} we conclude.

\paragraph{Supplementary materials.}
The full derivations can be found at
$$\texttt{\url{https://bitbucket.org/pl-uwr/scbv-machine}}$$






\section{Deconstruction of the KN Machine}
\label{sec:deconstruction}


In this section we highlight the endpoints of the derivation: the KN
machine, and the resulting evaluator obtained from an OCaml encoding
of the machine. The main steps in the derivation are: disentangling
the abstract machine into a defunctionalized form, refunctionalizing
the stacks of the machine into continuations, mapping the
continuation-passing evaluator to direct style, and refunctionalizing
the closures the direct-style evaluator operates on into their
functional representation~\cite{Danvy:IFL04}. All these
transformations are described in detail in the supplementary
materials.

\subsection{Specification of the KN Machine}

\begin{figure}[t!!]
\textbf{Syntax:}\\[-2em]
\begingroup
\begin{align*}
\mathit{Terms} \ni    T   &::= \tvar{n} \alt \tappp{T_1}{T_2} \alt \tolam T \\
T_N &::= T \alt \lvar{n} \\
\mathit{Closures} \ni C   &::= \clos{T_N} E  \\
\mathit{Envs} = \mathit{Closures}^* \ni E   &::= \nil \alt \cons{C}{E} \\
\mathit{Frames} \ni F &::= \frclos{T}{E} \alt \flam \alt \flstrong{T}\\
\mathit{Stacks} = \mathit{Frames}^* \ni S   &::= \nil \alt \cons{F}{S}\\
\mathit{Confs} \ni K   &::= \econf{T_N} E m S \alt \knsconf T m S 
\end{align*}
\endgroup
\textbf{Initial state (for closed terms):}
\begin{align*} I_{\mathit{KN}} : T \mapsto \econf T \nil 0 \nil\\[-2em]
\end{align*}
%
\textbf{Transition rules:}
\begingroup
\begin{align}
\econf {\tappp{T_1}{T_2}} E m {S_1} &\to \econf {T_1} E m {\cons{\frclos {T_2} E} {S_1}}\\
\econf {\tolam T} E m {\cons{\frclos{T'}{E'}}{S_1}} &\to \econf T {\cons{\clos{T'}{E'}}{E}} m {S_1}\\
\econf {\tolam T} E m {S_2} &\to \econf T {\cons{\clos {\lvar{m+1}} \nil} E} {m+1} {\cons{\flam}{S_2}}\\
\econf {\tvar{0}} {\cons{\clos T E}{E'}} m {S_1} &\to \econf T E m {S_1}\\
\econf {\tvar{n+1}} {\cons{C}{E}} m {S_1} &\to \econf {\tvar{n}} E m {S_1}\\
\econf {\lvar{n}} E m {S_1} &\to \knsconf {\tvar{m-n}} m {S_1}\\
\knsconf \nf 0 \nil &\to \nf \\
\knsconf \neu m {\cons{\frclos{T'}{E'}}{S_1}} &\to \econf {T'} {E'} m {\cons{\flstrong{\neu}}{S_1}} \\
\knsconf \nf m {\cons{\flam}{S_2}} &\to \knsconf {\tolam \nf} {m-1} {S_2} \\
\knsconf \nf m {\cons{\flstrong{\neu}}{S_1}} &\to \knsconf {\tappp{\neu}{\nf}} m {S_1}
\end{align}
\endgroup
\caption{Rules for the KN machine}
\label{fig:KN2020}
\end{figure}

Crégut's KN machine is shown in Figure~\ref{fig:KN2020}.
Due to the lack of space we do not discuss its architecture
here; we refer the reader to the original paper~\cite{Cregut:HOSC07}
(which also includes a nice introduction to de Bruijn indices and
levels) or to a more modern presentation
in~\cite{DBLP:journals/jfp/Garcia-PerezN19}. We also discuss all
transitions of the machine in our supplementary materials.

The presentation here is slightly optimized compared to the original,
and it coincides (on closed terms) with later presentation introduced
by Munk~\cite{Munk:MS}. The machine is in strong bisimulation with the
original one, but the latter threads more redundant information: the
parameter $m$ in configurations is exactly the number of lambda frames
in the current stack and need not be saved in stack frames.

The machine operates on lambda terms with de Bruijn indices used to
represent bound variables in the standard way. Things get more
complicated when we want to reduce open terms or reduce under lambdas,
where we need to care for free variables. In the KN machine this is
done using de Bruijn levels which represent the number of enclosing
lambda abstractions from the root of the term to the current variable
occurrence, and such \textit{abstract variables} are formed with a
different constructor $\lvar{n}$. The machine normalizes terms
according to the normal-order strategy that extends CbN to reduce open
terms and under lambdas. It can be seen as an extension of the Krivine
machine for CbN~\cite{Krivine:HOSC07}.



\subsection{Shape Invariant}
\label{sec:inv}
The machine specification can be seen as a function explicitly written
in trampolined style \cite{Ganz-al:ICFP99}, where each transition
dispatches by a single pattern matching on the term or on the stack
component of the configuration. Stacks are sequences of
frames that are constructed when traversing the term in search of a
next redex. However, this ``flat'' representation allows more stacks
to be formed than are reachable in a machine run from the initial
empty stack. In order to reason about the machine correctness, one
needs to identify the precise structure of reachable stacks.
Crégut expresses this shape invariant by a regular expression
\cite{Cregut:HOSC07} but it can also be expressed simply by the
context grammar using two kinds of stacks $S_1, S_2$:
\begin{align*}
S_1   &::= \cons{\frclos{T}{E}}{S_1} \alt S_2 \\
S_2 &::= \nil \alt \cons{\flam}{S_2} \alt \cons{\flstrong{\neu}}{S_1}
\end{align*}
\noindent
where $\neu$ denotes terms in neutral form. Neutral and normal forms
are constructed according to the following grammar:
\begin{align*}
\text{Normal forms} \ni \nf &::= \tolam {\nf} \alt \neu\\
\text{Neutral terms} \ni \neu &::= \tvar{n} \alt \tappp{\neu}{\nf}
\end{align*}
%

Garc{\'\i}a-P{\'e}rez \& Nogueira~\cite{Garcia-Nogueira:SCP14}
underline the importance of establishing the shape invariant for
refunctionalization step of the functional correspondence and
characterize evaluation contexts of the normal-order strategy by an
outside-in context grammar.\footnote{This family can be also defined
  in terms of an order on contexts~\cite{Accattoli-DalLago:LMCS16}.}
Below we present the grammar of normal-order contexts for the
$\lambda$-calculus, \ie,~leftmost-outermost contexts.  We can see that
the machine stacks correspond to the inside-out representation of
contexts: $S_1$ represents $L$-contexts encoding the weak CbN strategy
while $S_2$ represents $A$-contexts of the strong extension of CbN.
The grammar of outside-in contexts is, on the other hand, more natural
for top-down decomposition.  Both $L^{io}$ and $L^{oi}$ represent the
same $L$-contexts family but with reversed order of frames in the
lists.  We elaborate on the connection between the two kinds of
representations in Section~\ref{sec:reduction-semantics} when we
discuss the strong CbV strategy.

\begin{multicols}{2}
\begin{center}
inside-out contexts
\begin{align*}
L^{io} &::= \cons{\frapp T}{L^{io}} \alt A^{io}\\
A^{io} &::= \ctxnil \alt \cons{\lambda\hole}{A^{io}} \alt \cons{\flapp {\neu}}{L^{io}}
\end{align*}

\end{center}
\columnbreak
\begin{center}
outside-in contexts
\begin{align*}
B^{oi} &::= \cons{\frapp {T}}{B^{oi}} \alt \cons{\flapp {\neu}}{L^{oi}} \alt \ctxnil \\
L^{oi} &::= \cons{\lambda\hole}{L^{oi}} \alt B^{oi}
\end{align*}

\end{center}
\end{multicols}


\subsection{Compositional Evaluator}

\begin{figure}[t!!]
  \centering
  \begin{lstlisting}
(* syntax of the lambda-calculus with de Bruijn indices *)
type index = int
type term  = Var of index | Lam of term | App of term * term

(* semantic domain *)
type level = int
type glue  = Abs of (sem -> sem) | Neutral of term
 and sem   = level -> glue

(* reification of semantic objects into normal forms *)
let rec reify (d : sem) (m : level) : term =
  match d m with
  | Abs f ->
    Lam (reify (f (fun m' -> Neutral (Var (m'-m-1))))(m+1))
  | Neutral a ->
    a

(* sem -> sem as a retract of sem *)
let to_sem (f : sem -> sem) : sem =
  fun _ -> Abs f
let from_sem (d : sem) : sem -> sem =
  fun d' -> fun m ->
    match d m with
    | Abs f -> f d' m
    | Neutral a -> Neutral (App (a, reify d' m))

(* interpretation function *)
let rec eval (t : term) (e : sem list) : sem =
  match t with
  | Var n -> List.nth e n
  | Lam t' -> to_sem (fun d -> eval t' (d :: e))
  | App (t1, t2) -> from_sem (eval t1 e)
                             (fun m -> eval t2 e m)

(* NbE: interpretation followed by reification *)
let nbe (t : term) : term = reify (eval t []) 0
  \end{lstlisting}
  \caption{An OCaml implementation of the higher-order compositional
    evaluator corresponding to the KN machine: an instance of
    normalization by evaluation for normal-order $\beta$-reduction in
    the $\lambda$-calculus.}
  \label{fig:nbe-cbn}
\end{figure}

The evaluator derived through the functional correspondence from the
encoding of the abstract machine of Figure~\ref{fig:KN2020}, after
some tidying to underline its structure, is shown in
Figure~\ref{fig:nbe-cbn}. The evaluator implements an algorithm that
follows the principles of normalization by
evaluation~\cite{Filinski-Rohde:RAIRO05}, where the idea is to map a
$\lambda$-term to an object in the meta-language from which a
syntactic normal form of the input term can subsequently be read
off. Actually, what we have mechanically obtained from KN is an OCaml
implementation of a domain-theoretic residualizing model of the
$\lambda$-calculus, in which the recursive type {\tt sem} is an
encoding of a reflexive domain $D$ of interpretation, isomorphic to
$\mathcal{N} \rightarrow ((D \rightarrow D) +
\Lambda^{neu}_\bot)_\bot$ (where $\mathcal{N}$ and $\Lambda^{neu}$ are
discrete CPOs of natural numbers and neutral terms, respectively). In
particular, {\tt to\_sem} and {\tt from\_sem} encode continuous
functions $\phi : (D \rightarrow D) \rightarrow D$ and
$\psi : D \rightarrow (D \rightarrow D)$, respectively, such that
$\psi \circ \phi = id$, establishing that $D \rightarrow D$ is a
retract of $D$~\cite{Filinski-Rohde:RAIRO05}, which guarantees that
$\beta$-convertible terms are mapped to the same semantic object. The
interpretation function {\tt eval} is completely standard, except for
the $\eta$-expansion in the clause for application which comes from
the fact that the derivation has been carried out in an eager
meta-language. The reification function {\tt reify} mediates between
syntax and semantics in the way known from Filinski and Rohde's
work~\cite{Filinski-Rohde:RAIRO05} on NbE for the untyped
$\lambda$-calculus.

As a matter of fact, what we have obtained through the functional
correspondence from KN is very close to what Filinski and Rohde
invented (and proved correct using domain-theoretic tools). The
difference lies in the semantic domain which in their case was
represented (in SML) by the type that in OCaml would read as
\begin{lstlisting}
type sem = Abs of ((unit -> sem) -> sem)
         | Neutral of (level -> term)
\end{lstlisting}
from which we can see that the de Bruijn level is only needed to
construct a neutral term and otherwise redundant (an observation
confirmed by the definition of {\tt to\_sem} we have derived). With
this domain of interpretation function arguments are explicitly passed
as thunks. From the reduction strategy point of view, the normalizer
of Figure~\ref{fig:nbe-cbn} (and KN) implements a two-stage
normalization strategy: first reduce a term to a weak normal form
(function {\tt eval}) and then normalize the result (function {\tt
  reify}). Seen that way, the two constructors of type {\tt sem}
represent the two possible kinds of weak normal forms.

For the record, we have derived an alternative abstract machine for
normal-order reduction starting with Filinski and Rohde's NbE. This
machine differs from KN in that it processes neutral terms in a
separate mode and with an additional kind of stack. In the next
section, we modify our NbE so that it accounts for CbV function
applications.

In his MSc thesis, Munk also presents selected steps of a
deconstruction of KN into a NbE~\cite{Munk:MS}. However, he goes
through a step in which de Bruijn levels are moved from the stack to
closures in the environment. This step has not been formally justified
and the resulting NbE is quite different from Filinski and Rohde's or
from ours.


\section{Construction of a Call-by-Value Variant}
In this section we derive a call-by-value variant of the Crégut
abstract machine. This is done by modifying the evaluator of
Figure~\ref{fig:nbe-cbn} such that it accounts for CbV, and then
inverting the transformations on the path from the abstract machine to
the evaluator.

Call by value is a family of strategies where arguments of a function
are evaluated (to a weak normal form) before being passed to the
function. This way one avoids needless recomputation of arguments that
are used more than once. 
A possible approach to a strong variant of such a strategy is the
\emph{applicative order} (a.k.a. leftmost-innermost)
reduction~\cite{Sestoft:Jones02}, where the arguments are evaluated to
the strong normal form.
Here, however, we aim at a different, two-stage strategy, analogous to
the one embodied in KN and in the normalizer of
Figure~\ref{fig:nbe-cbn}, which is a conservative extension of the
standard CbV: the arguments are first evaluated to a weak normal form,
then the function is applied and only then the resulting weak normal
form is further reduced to the strong normal form. In order to obtain
one fixed member of the family, we
follow~\cite{DBLP:journals/scp/AccattoliG19} and choose the
right-to-left order of evaluation of arguments (we also choose the
right-to-left order of normalization in inert terms, see
Section~\ref{sec:reduction-semantics}).



\label{sec:cbv-machine}

\begin{figure}[t!!]
  \centering
  \begin{lstlisting}
type sem = Abs of (sem -> sem) | Neutral of (level -> term)

let rec reify (d : sem) (m : level) : term =
  match d with
  | Abs f ->
    Lam (reify (f (Neutral (fun m' -> Var (m'-m-1))))(m+1))
  | Neutral l ->
    l m

let to_sem (f : sem -> sem) : sem = Abs f

let from_sem (d : sem) : sem -> sem =
  fun d' ->
    match d with
    | Abs f ->
      f d'
    | Neutral l ->
      Neutral (fun m -> let n = reify d' m in App (l m, n))

let rec eval (t : term) (e : sem list) : sem =
  match t with
  | Var n -> List.nth e n
  | Lam t' -> to_sem (fun d -> eval t' (d :: e))
  | App (t1, t2) -> let d2 = eval t2 e
                    in from_sem (eval t1 e) d2

let nbe (t : term) : term = reify (eval t []) 0
  \end{lstlisting}
  \caption{An OCaml implementation of the modified higher-order
    compositional evaluator: an instance of normalization by
    evaluation for a call-by-value $\beta$-reduction in the
    $\lambda$-calculus.}
  \label{fig:nbe-cbv}
\end{figure}

\subsection{Call-by-Value Evaluator}
In call by value, function arguments are evaluated before the
application takes place. To reflect this design choice in the
evaluator, we modify the domain of interpretation:
\begin{lstlisting}
type sem = Abs of (sem -> sem)
         | Neutral of (level -> term)
\end{lstlisting}
where an argument passed to a function is no longer a thunk, but a
preevaluated value in the semantic domain. Here, the two constructors
correspond to two kinds of weak normal forms: $\lambda$-abstraction
and inert term, as presented in
Section~\ref{sec:reduction-semantics}.  All the other changes in the
evaluator are simple adjustments to this modification. An OCaml
implementation of the modified evaluator is shown in
Figure~\ref{fig:nbe-cbv}, where we arbitrarily decided to evaluate
function application right to left (witness the explicit sequencing of
computations with {\tt let} in the clause for application in {\tt
  eval}) and similarly for generating neutral terms (again, with {\tt
  let} in {\tt from\_sem}).~\footnote{While it would be possible to
  directly use Filinski and Rohde's NbE to obtain the evaluator of
  this section, our goal was to reveal and adjust the evaluator
  underlying KN, and the precise relation between KN and Filinski and
  Rohde's NbE has not been revealed prior to this work.}

This normalizer could subsequently be given a domain-theoretic
treatment, using the same techniques as the ones applied by
Filinski and Rohde to their call-by-name
normalizer~\cite{Filinski-Rohde:RAIRO05} -- an interesting endeavour
that would offer one possible way of revealing the precise meaning of
the modified normalizer. Here, instead, we take advantage of the
functional correspondence and we derive a semantically equivalent
abstract machine that we then analyse and we identify the reduction
strategy it implements and inherits from the underlying NbE
of~Figure~\ref{fig:nbe-cbv}.

The machine we derived from the evaluator has been subject to further
optimizations before we arrived at the version presented in the next
section. In particular, the de Bruijn level $m$ has been moved from
application frames of the stack to a dedicated register in the
configurations of the machine. This modification requires a more
careful bookkeeping of the level and, most notably, it has to be
decremented when the machine leaves the scope of a lambda, just as in
KN of Figure~\ref{fig:KN2020}. We also flattened the stack structure
to be represented by a single list of frames, instead of by a pair of
mutually inductive list-like structures. The final machine is then
close in style to KN and can be seen as its call-by-value
variant.


\subsection{Abstract Machine}

\begin{figure}[t]
\textbf{Syntax:}
\begin{align*}
\mathit{Terms}  \ni T   &::= \tvar{n} \alt \tappp{T_1}{T_2} \alt \tolam T \\
\mathit{Wnfs}   \ni W   &::= \abstr T E \alt I\\
\mathit{Inerts} \ni I   &::= \lvar{n} \alt \tappp I W\\
\mathit{Envs}   \ni E   &::= \nil \alt \cons{W}{E} \\
\mathit{Frames} \ni F   &::= \flclos{T}{E}
    \alt \frapp W
    \alt \frstrong{T} 
    \alt \flam
    \alt \flstrong{I} \\
\mathit{Stacks} \ni S   &::= \nil \alt \cons{F}{S} \\
\mathit{Confs} \ni K   &::=
  \wconf T E S m
  \alt \bconf W S 
  \alt \sconf T m S
\end{align*}
%
%
%
%
\setcounter{equation}{0}
\textbf{Transition rules:}
\begin{align}
  \setcounter{equation}{-1}
   T &\mapsto \wconf T {\nil} {\nil} 0 \label{tr:0}\\
\wconf {\tappp{T_1}{T_2}} E {S_1} m &\to \wconf {T_2} E {\cons{\flclos{T_1} E} {S_1}} m \label{tr:1}\\
\wconf {\tolam T} E {S_1} m &\to \bconf {\abstr T E} {S_1}\label{tr:2}\\
\wconf {\tvar 0} {\cons W E} {S_1} m &\to \bconf W {S_1}\label{tr:3}\\
\wconf {\tvar{n+1}} {\cons W E} {S_1} m &\to \wconf n E {S_1} m\label{tr:4}\\
\bconf W {\cons {\flclos T E} {S_1}} &\to \wconf T E {\cons{ \frapp W} {S_1}} m\label{tr:5}\\
\bconf {\abstr T E} {\cons {\frapp W} {S_1}} &\to \wconf T {\cons W E} {S_1} m\label{tr:6}\\
\bconf I {\cons {\frapp W} {S_1}} &\to \bconf {\tappp I W} {S_1}\label{tr:7}\\
\bconf {\abstr T E} {{S_3}} &\to \wconf T {\cons {\lvar{m+1}} E} {\cons \flam {S_3}} {m+1}\label{tr:8}\\
\bconf {\tappp I W} {{S_2}} &\to \bconf W {{\cons {\flstrong I} {S_2}}}\label{tr:9}\\
\bconf {\lvar n} {{S_2}} &\to \sconf {m-n} m {S_2}\label{tr:10}\\
\sconf \nf m {\cons {\flstrong I} {S_2}} &\to \bconf I {{\cons{\frstrong \nf} {S_2}}}\label{tr:13}\\
\sconf \nf m {\cons \flam {S_3}} &\to \sconf {\tolam \nf} {m-1} {S_3} \label{tr:12}\\
\sconf {\neu} m {\cons{\frstrong {\nf}} {S_2}} &\to \sconf {\tappp {\neu} {\nf}} m {S_2} \label{tr:14}\\
\sconf \nf 0 \nil &\mapsto \nf \label{tr:11}
\end{align}
%
\caption{Rules for KNV, a call-by-value variant of KN}\label{KNV}
\end{figure}

The machine obtained by derivation from the NbE evaluator is presented
in Figure~\ref{KNV}. There are syntactic categories of lambda terms
$T$ in de Bruijn notation, machine representations of weak normal
forms $W$, inert terms $I$, environments~$E$, stack frames $F$, stacks
$S$ and configurations $K$.  Weak normal forms are either closures
consisting of a lambda abstraction and an environment or inert
terms. Inert terms are either abstract variables $V(n)$ or inert terms
applied to a weak normal form. Just as in the KN machine, here $n$ is de
Bruijn level (not to be confused with de Bruijn index in the grammar
of terms $T$). Weak normal forms represent the intermediate values
that are passed to functions as arguments through the environment and
subsequently reduced further to normal form. Environments are just
sequences of weak normal forms; they represent mappings that assign
$n$th element of the sequence to the variable with de Bruijn index
$n$.  As usual, stacks represent evaluation contexts.


There are three kinds of configurations corresponding to three modes
of operation: in configurations $\wconf \cdot{\cdot}\cdot{\cdot}$ the
machine evaluates some subterm to a weak normal form; in
$\bconfwiththreearguments \cdot \cdot \cdot$ it continues with a
computed weak normal form and in $\sconf \cdot {\cdot}\cdot$ it
continues with a computed (strong) normal form.  Let us discuss the
transitions. For the moment we ignore the indices in stacks; we think
of $S_1, S_2$ and $S_3$ as arbitrary members of the syntactic category
$S$ of stacks. Similarly, $\neu$ and $\nf$ are arbitrary terms. These
indices will become relevant in the next section.

Transitions (\ref{tr:0})--(\ref{tr:6}) implement a right-to-left
version of the well-known CEK machine~\cite{Felleisen-Friedman:FDPC3}
in a~formulation similar to \cite{Ager-al:PPDP03}, but using de Bruijn
indices. The initial transition~(\ref{tr:0}) loads the term to be
evaluated to a configuration with the empty environment and empty
stack on de Bruijn level 0.  Transitions (\ref{tr:1})--(\ref{tr:4})
operate on configurations of the form $\wconf T E {S} m$ that are
meant to evaluate the term $T$ within the environment $E$ in the
context represented by $S$ to a~weak normal form (wnf). In the case of
application $\tappp{T_1}{T_2}$, transition (\ref{tr:1}) calls the
evaluation of $T_2$ and pushes a closure pairing $T_1$ with the
current environment to the stack. Note that this implements the first
of our right-to-left choices of the order of reduction. A~lambda
abstraction in (\ref{tr:2}) is already in wnf, so we change the mode
of operation to a~configuration of the form $\bconf W
{S}$. Transitions (\ref{tr:3}) and~(\ref{tr:4}) simply read a~value of
a~variable from the environment (which always returns a~wnf) and
change the mode of operation.

Configurations of the form $\bconf W {S}$ continue with a~wnf $W$ in
a~context represented by $S$. There are two goals in these
configurations: the first is to finish the evaluation (to wnfs) of the
closures stored on the stack according to the weak call-by-value
strategy; the second is to reduce $W$ to a~strong normal form. This is
handled by rules (\ref{tr:5})--(\ref{tr:10}), where rules (\ref{tr:5})
and (\ref{tr:6}) are responsible for the first goal, and rules
(\ref{tr:7})--(\ref{tr:10}) for the second. In rule (\ref{tr:5}) the
stack contains a~closure, so we start evaluating this closure and push
the already computed wnf to the stack; when this evaluation reaches
a~wnf, rules (\ref{tr:6}) or (\ref{tr:7}) apply. If the wnf is
a~lambda abstraction, transition (\ref{tr:6}) implements
a~$\beta$-contraction. Otherwise it is an inert term; in this case
rule (\ref{tr:7}) reconstructs the application of this inert term to
the wnf popped from the stack (which gives another wnf). Rules
(\ref{tr:8})--(\ref{tr:10}) are applied when there are no more wnfs on
the top of the stack; here we pattern-match on the currently processed
wnf $W$. If $W$ is a~closure, transition (\ref{tr:8}) pushes the
elementary context $\flam$ to the stack, increments the de Bruijn
level ($m+1$), adds the abstract variable $\lvar n$ to the environment
and starts the evaluation of the body. If $W$ is an application
$\tappp{I}{W'}$, rule (\ref{tr:9}) delays the normalization of $I$ by
pushing it to the stack and continues with $W'$; note that this
implements the second of our right-to-left choices of the order of
reduction. Finally, if $W$ is an abstract variable with index $n$ at
level $m$, we reach a~normal form; rule (\ref{tr:10}) computes the
final index of the variable and changes the mode of operation to
a~configuration of the form $\sconf T m {S}$.

Configurations of the form $\sconf T m {S}$ continue with a~(strong)
normal form~$T$ in a~context represented by $S$ (recall that the
grammar of normal forms is presented in Section~\ref{sec:inv}). The
goal in these configurations is to finish the evaluation of inert term
stored on the stack and to reconstruct the final term. This is handled
by transitions (\ref{tr:13})--(\ref{tr:11}); the choice of the
transition is done by pattern-matching on the stack. If there is an
inert term $I$ on the top of the stack, rule (\ref{tr:13}) pushes the
already computed normal form on the stack and calls normalization of
$I$ by switching the mode of operation to $\bconf W {S}$. Otherwise
there is a~$\flam$ frame or a~previously computed normal form on the
top of the stack; in these cases transitions (\ref{tr:12}) and
(\ref{tr:14}) reconstruct the term accordingly. Finally, when the
stack is empty, transition (\ref{tr:11}) unloads the final result from
a~configuration.


The machine is pure in a sense that it does not use mutable state nor
other computational effects so it can be directly implemented in a pure
functional language. Thanks to that all structures of the machine are
persistent data structures with their advantages (cf.~\cite{Okasaki:99}).
It differs from machines of \cite{DBLP:journals/scp/AccattoliG19}
in that its implementation does not perform on-the-fly $\alpha$-conversion
nor does it use pointers explicitly.
Assuming \textit{uniform cost criteria} for arithmetic operations,
the cost of dispatch and of each transition is constant.


\subsection{Shape invariants}

As in the case of the KN machine, not all sequences of stack frames
represent valid contexts that can occur in a reachable configuration
of the machine from Figure~\ref{KNV}. We define the syntactic category
$\kwf$ of well-formed configurations with the following grammar.
%
\begin{align*}
S_1 &::=
       \cons{\flclos{T}{E}}{S_1}
  \alt \cons{\frapp W}{S_1}
  \alt {S_3}\\
S_2 &::= \cons{\frstrong{\nf}}{S_2} \alt S_3\\
S_3 &::= \nil
  \alt \cons{\flam}{S_3}
     \alt \cons{\flstrong{I}}{S_2} \\
\kwf&::=
  \wconf T E {S_1} m
  \!\!\alt\! \bconf W {S_1}
  \!\!\alt\! \bconf I {S_2}  
  \!\!\alt\! \sconf \neu m {S_2}
  \!\!\alt\! \sconf \nf m {S_3}
\end{align*}
A simple induction on the length of evaluation gives the folowing lemma. 
\begin{lemma}\label{lem:wellformed}
  For all initial terms $T$, all configurations reachable from $T$ are
  well-formed.
\end{lemma}
One can note that there is no invention in designing
syntactic categories {$W$~and~$I$} which correspond to grammars
of weak normal forms and inert terms. They are products of
defunctionalization which is a part of mechanization
carried out via functional correspondence.
More interestingly, all shape invariants can be derived.  It is enough
to use a separate grammar for normal forms in the higher-order
normalizer.  Through derivation this grammar is imprinted on the
grammars of stacks and configurations.


\subsection{An Application: Streaming of Expressions}

Here we show a method for early discovering that two terms are not
$\beta$-convertible, i.e., that they do not have equal normal forms. 
Gr\'egoire\&Leroy show in \cite{Gregoire-Leroy:ICFP02} that the
comparison of normal forms can be short-circuited when enough data is
computed. Our idea is to run the machine on both terms as long as
partial results are the same. If the machine completes the computation
on both terms and the computed normal forms are equal, the terms are
$\beta$-convertible.  But whenever it sees partial results that are
different for the two input terms, we immediately know that the two
terms do not have equal normal forms, without actually completing the
computation.
In some cases it allows to give an answer even on divergent terms.  To
get a partial result it is enough to interrupt the machine after
transitions (\ref{tr:8}) and (\ref{tr:13}) when it pushes $\flam$ and
$\frstrong \nf$ frames, respectively, on the stack. This method is
implemented
in the accompanying code. 

As an example, consider the terms $\tlam{x}{\tlam{y}{\Omega}}$ and
$\tlam{x}{\tappp{(\tappp{x}{\tlam{y}{\Omega}})}{x}}$
(using standard notation with names).  Even if the
evaluation of these terms never terminates, we can detect different
partial results and determine that these two terms cannot have equal
normal forms. By running our machine we learn that the normal form of
the former term (if it exists) starts with
$\tlam{x}{\tlam{y}{\hole}}$ while the the normal form of the latter
(if it exists) starts with $\tlam{x}{\tappp{\hole}{x}}$.

This application is not specific to KNV and a similar procedure can be
implemented based on KN. However, as is usual with CbV vs CbN, KNV
in general performs better by avoiding reevaluation (to weak normal form)
of function arguments.



\section{Reduction semantics for strong CbV}
\label{sec:reduction-semantics}
\begin{figure}[t!!]
\begin{multicols}{2}
\begin{center}
\begin{align*}
\indsub i T {(\tappp {T_1} {T_2})} &= \tappp {\indsub i T {T_1}}
        {\indsub i T {T_2}}\\ \indsub i T {(\tolam {T_1})} &= \tolam
        {(\indsub {i+1} T {T_1})}\\ \indsub i T {n} &= \begin{cases} n
          &: n < i\\ \indsft i 0 T &: n = i\\ n-1 \;\; &: n >
          i \end{cases}
\end{align*}
\end{center}
\columnbreak
\begin{center}
\begin{align*}
\indsft i k {(\tappp {T_1} {T_2})} &= \tappp {\indsft i k {T_1}} {\indsft i k {T_2}}\\
\indsft i k {\tolam T}             &= \tolam {\indsft i {k+1} T}\\
\indsft i k n        &= \begin{cases} n + i\;\; &: n \geq k\\
                                      n         &: n < k \end{cases}
\end{align*}
\end{center}
\end{multicols}
\vspace{-3em}
\begin{center}
\begin{align*}
(\lambda T_1) T_2 & \rightharpoonup_\beta \indsub 0 {T_2} {T_1} \\
\phantom{
\;\;\;\; \text{if} \;\; T_1 \rightharpoonup_\beta T_2} 
C[T_1] & \stackrel C {\to}_\beta C[T_2]
\;\;\;\; \text{if} \;\; T_1 \rightharpoonup_\beta T_2 
\end{align*}
\end{center}
\caption{$\beta$-contraction and $\beta$-reduction for terms with de Bruijn indices}
\label{fig:beta}
\end{figure}

A reduction semantics is a form of small-step operational semantics
with an explicit representation of reduction contexts, \ie, of
locations in a term where the computation can take place. Roughly,
reduction contexts can be thought to represent terms with a hole. The
atomic computation step is defined by a rewriting relation on terms,
often called the contraction relation. For example, in our source
language, the lambda calculus with de Bruijn indices, the reduction
semantics can be formally defined as in Figure~\ref{fig:beta}, where
the contraction relation is $\rightharpoonup_{\beta}$, and one-step
reduction is defined as contraction in context $\stackrel C
{\to}_\beta$, where reduction contexts $C$ describe the specific
reduction strategy. For example, if we take $C$ to be $L$ from
Section~\ref{sec:inv}, then we obtain the normal-order strategy. The
notation $C[T]$ denotes the term reconstructed by {\em plugging} the
term $T$ in the hole of the context~$C$.

In uniform strategies the grammar of reduction contexts is defined
using only one nonterminal (as in CbN or CbV), while {\it hybrid}
strategies use more than one nonterminal (as in normal order). The
strong CbV strategy is another example of a hybrid strategy, one with
three nonterminals leading to three kinds of contexts, each describing
a separate substrategy.

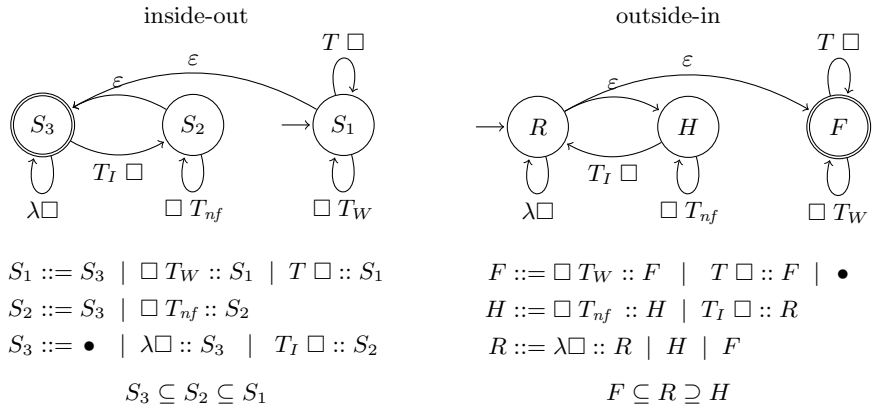
\begin{figure}[htb]
\begin{multicols}{2}
\begin{center}
inside-out
\begin{tikzpicture}[shorten >=1pt,node distance=2cm,on grid,auto, initial text={}]
\node[state,accepting]   (N)              {$S_3$};
\node[state]           (P) [right of=N] {$S_2$};
\node[state,initial] (C) [right of=P] {$S_1$};

\path[->]
(N) edge [loop below] node {$\lambda\hole$}   ( )
(P) edge [bend right, above] node {$\varepsilon$}    (N)
(C) edge [bend right, above] node {$\varepsilon$}    (N)
(P) edge [loop below] node {$\frapp {\nf}$} ()
(N) edge [bend right, below] node {$\flapp {\inert}$} (P)
(C) edge [loop below] node {$\frapp \wnf$} ()
    edge [loop above] node {$\flapp T$} ();
\end{tikzpicture}
\begin{align*}
S_1 &::= S_3 \alt \cons {\frapp \wnf} {S_1} \alt \cons{\flapp T} {S_1}\\
S_2 &::= S_3 \alt \cons {\frapp {\nf}}{S_2}\\
S_3 &::= \ctxnil \hspace{1mm} \alt \cons{\lambda\hole}{S_3} \hspace{1mm} \alt \hspace{1mm} \cons{\flapp {\inert}}{S_2}
\end{align*}

$S_3 \subseteq S_2 \subseteq S_1 $

\end{center}
\columnbreak
\begin{center}
outside-in
\begin{tikzpicture}[shorten >=1pt,node distance=2cm,on grid,auto, initial text={}]
\node[state,initial]   (N)              {$R$};
\node[state]           (P) [right of=N] {$H$};
\node[state,accepting] (C) [right of=P] {$F$};

\path[->]
(N) edge [loop below] node {$\lambda\hole$}   ( )
    edge [bend left]  node {$\varepsilon$}    (P)
    edge [bend left]  node {$\varepsilon$}    (C)
(P) edge [loop below] node {$\frapp {\nf}$} ()
    edge [bend left]  node {$\flapp {\inert}$} (N)
(C) edge [loop below] node {$\frapp \wnf$} ()
    edge [loop above] node {$\flapp T$} ();
\end{tikzpicture}\vspace{-2ex}%
\begin{align*}
F &::= \cons{\frapp \wnf} F \hspace{0.8mm}\alt \hspace{1.3mm}\cons{\flapp T} F \alt \ctxnil\\
H &::= \cons{\frapp {\nf} \hspace{1mm}} H \alt \cons{\flapp {\inert}} R \\
R &::= \cons{\lambda\hole} R \alt H \alt F\\[-2.4em]
\end{align*}
$ F \subseteq R \supseteq H $
\end{center}
\end{multicols}
\caption{Reduction semantics: automata and grammars of contexts}
\label{fig:automata}
\end{figure}

As already observed in Section~\ref{sec:inv}, shape invariants of the machine stack
naturally lead to
reduction contexts of the strategy realized by the machine. For the
case of the KNV
machine, stack invariants translate to 
grammar of contexts shown in Figure~\ref{fig:automata}
(left). Equivalently, they can be translated to an automaton, whose
transitions
are labelled with terms (as opposed to their machine
representations), where the syntactic categories of terms in weak
normal form and inert terms in the lambda calculus are
\begin{align*}
\text{Weak normal forms } \setofwnf \ni \wnf &::= \tolam {T} \alt \inert\\
\text{Inert terms } \setofinert \ni \inert &::= \tvar{n} \alt \tappp{\inert}{\wnf}
\end{align*}
The grammar generates all stacks in syntactic categories $S_1,S_2,S_3$
in an inside-out manner: the automaton reading a stack from left to
right moves from the hole of the represented context towards the
topmost symbol. By reversing the arrows in the automaton we obtain an
outside-in grammar (Figure~\ref{fig:automata} right).  A context
nonterminal (its kind) in inside-out grammars denotes the kind of the
hole, whereas in outside-in grammars it denotes the kind of the
context generated by that nonterminal.

To complete the reduction semantics of strong CbV we have to
specify a contraction relation. We simply read it from transition
(\ref{tr:6}), where environments storing delayed substitution
consist of structures representing weak normal forms. Thus our
contraction is $\beta$-contraction restricted to a variant where an
argument has to be in weak normal form.  We call it
$\bwnf$-contraction:
$$(\lambda T) \wnf \rightharpoonup_{\beta{wnf}} \indsub 0 {\wnf} {T}$$


%




The substrategy corresponding to the $F$ nonterminal in
Figure~\ref{fig:automata} and $\bwnf$-contraction can be recognized as
the right-to-left weak strategy of the \textit{fireball calculus}
considered in~\cite{DBLP:journals/scp/AccattoliG19}. It is known that
this strategy is deterministic and reduces terms to weak normal
forms. Our strong strategy corresponds to the nonterminal $R$ (the
starting symbol in the grammar); it contains the substrategy~$F$ and
thus it is a conservative extension of the right-to-left call-by-value
strategy.

In our strategy arguments of functions are
evaluated in the right-to-left order. Similarly, arguments of inert
terms are evaluated in the same order---thus we can refer to the
strategy as \emph{twice right-to-left call-by-value, rrCbV}. This is
an arbitrary choice; three other options are possible. Some of these
options, like lrCbV leave place for optimizations: after completing
the weak right-to-left reduction the stack contains a sequence of
arguments in weak normal form, which are then composed to build an
inert term that is immediately decomposed to the very same sequence of
weak normal forms before normalizing them with the strong
left-to-right strategy. An optimized machine could refocus directly to
strong reduction of arguments on stack instead of rebuilding an inert
term and decomposing it again.

Strong CbV, as weak CbV, is an \textit{incomplete strategy},
\ie,~some normalizable terms may loop forever,
\eg,~$\tappp {\tappp K I} \Omega$.\footnote
{Where $K = \tlam x {\tlam y x}$, $I = \tlam x x$,
$\Omega = \tapp {(\tlam x {\tapp x x})}{(\tlam x {\tapp x x})}$,
using standard notation with names.}
Nevertheless, it allows to compute values of recursive functions.

\begin{example}
Consider the term $\tolam {(\tappp {\tappp K I} \Omega)}$.
We can decompose it uniquely
into a context $\tolam {(\tappp {\tappp K I} \hole)}$
and a subterm $\Omega$ forming a $\bwnf$-redex.
The context in the inside-out representation is
$\cons{\flapp {\tappp K I}} {\cons \flam \ctxnil}$
and it satisfies the $S_1$ constraints.
In the outside-in representation it is
$\cons\flam {\cons {\flapp {\tappp K I}} \ctxnil}$
and conforms with the grammar $R$.
Here $S_1$ and $R$ are initial nonterminals in the grammars
of contexts defined in Figure~\ref{fig:automata}.
Thus $\tolam {(\tappp {\tappp K I} \Omega)}$
loops in the rrCbV strategy.

On the contrary, in term
$\tolam {(\tappp {\tappp K I} {\tolam \Omega})}$
the subterm $\tolam \Omega$ fits the $\wnf$ grammar
and $\tolam {(\tappp {\hole} {\tolam \Omega})}$
is a correct context of rrCbV.
Thus $\tolam {(\tappp {\tappp K I} {\tolam \Omega})}$
reduces to $\tolam I$ in two steps.

\end{example}

\section{Correctness}
\label{sec:correctness}

In this section we show the correctness of the derived machine: it
\emph{traces}~\cite{BChZ-fscd17} (\ie, exactly implements, in a
step-by-step manner) the reduction semantics. Before stating the
formal theorem we need some definitions.
\subsection{Decoding of machine representations}
\subsubsection*{Terms.}
In the proof of correctness we have to translate lambda terms to
machine configurations and back. The encoding of a term to a
configuration is given by transition~(\ref{tr:0}) in
Figure~\ref{KNV}. The translation in the other direction is more
involved. We start by defining two functions:
$\semsub {\cdot, \cdot} W : \mathit{Wnfs} \to \mathbb N \to \setofwnf$
decoding the machine representations of weak normal forms and the
function
$\semsub {\cdot, \cdot} I : \mathit{Inerts} \to \mathbb N \to
\setofinert$ decoding the representations of inert terms.  The formal
definitions of these functions are given in Appendix~\ref{app:decoding}.
The second parameter, which is a de
Bruijn level, is needed to decode an abstract variable.  The function
$\semsub{\cdot, \cdot, \cdot} T : \textit{Terms} \to \textit{Envs} \to
\mathbb N \to \textit{Terms}$ decodes machine representations of
arbitrary terms.
\subsubsection*{Stacks.}
Intuitively, a stack should be decoded to an evaluation
context.  However, we are going to prove a termination result
(cf. Lemma~\ref{lem:termination}), for which we need an intermediate
representation: lists of \emph{annotated frames}. The annotation
$\strongWannotation{\cdot}$ in $\flstrongann{T}$ indicates that the
term $T$ occurring in a context $ \flapp T$ is known to be in weak
normal form; similarly $\strongNannotation{\cdot}$ in
$\frstrongann{T}$ indicates that $T$ is known to be in strong normal
form.
\begin{align*}
\mathit{AnnFrms} \ni C   &::= \flapp T
    \alt \frapp T
    \alt \flstrongann{T} 
    \alt \frstrongann{T} 
    \alt \flam\\
\mathit{AnnFrms}^* \ni L   &::= \nil \alt \cons{C}{L} 
\end{align*}
The function
$\semsub \cdot {S} : \mathit{Stacks} \to \mathit{AnnFrms}^*$ decodes
stacks by decoding term representations in stack frames and adding
frame annotations.

\subsubsection*{Annotated decompositions.}  A configuration of
the machine encodes, among others, a decomposition of a term into its subterm
and a surrounding context. 
\begin{example}\label{ex:evalomega}
  Consider a fragment of evaluation of the term $\lambda 00$ (which is
  $\lambda x. x x $ in de Bruijn notation). We adopt here the OCaml
  notation for lists, so ${\stacktwo{1}{\flam}}$ is the same as
  ${\stackconstwo{1}{\flam}}$.
\begin{align*}
\bconfone {V(1)} {{\stacktwo{\frapp{V(1)}}{\flam}}}
&\stackrel{(\ref{tr:7})}{\to}
\bconfone {\tappp{V(1)} {V(1)}} {\stackone{\flam}}
\stackrel{(\ref{tr:9})}{\to}\\
\bconfone {V(1)} {\stacktwo{\flstrong {V(1)}}{\flam}}
&\stackrel{(\ref{tr:10})}{\to}
\sconf 0 1 {\stacktwo{\flstrong {V(1)}}{\flam}}
\stackrel{(\ref{tr:13})}{\to}\\[2mm]
\bconfone {V(1)} {\stacktwo{\frstrong {0}}{\flam}}
\end{align*}
\end{example}
Here both stacks $\stacktwo{\frstrong {0}}{\flam}$ and
$\stacktwo{\frapp{V(1)}}{\flam}$ represent the same context
$\lambda(\frapp{0})$, so the first and the last configuration in this
sequence gives the same decomposition of $\lambda 00$ to the subterm
$0$ in the context $\lambda(\frapp{0})$.  In order to capture the fact
that the machine does not fall into an infinite loop, even if it
considers the same decomposition more than once, we introduce a more
informative notion of \emph{annotated decomposition}.  We introduce
annotations for terms that allow to distinguish between arbitrary
terms and terms in weak or strong normal form.
\begin{align*}
  \mathit{AnnTerms} \ni A   &::= T \alt \strongWannotation{T}\alt  \strongNannotation{T}\\
  \mathit{AnnDcmp} \ni D    &::= \cons A L
\end{align*}

\subsubsection*{Configurations.} Configurations are first decoded to
annotated decompositions with function
$\semsub \cdot K : \textit{Confs}\to \mathit{AnnDcmp}$ and then to terms by function
$\plg: \mathit{AnnDcmp} \to \mathit{Terms}$. The latter function ignores all annotations.

\subsection{Formal correctness result}

We are now ready to state the result formally as the following theorem. 
\begin{theorem}
\label{thm:correctness}
  KNV traces the twice right-to-left strong CbV strategy, \ie:
  \begin{enumerate}
  \item \label{cor:surj} The function
    $\plg(\semsub{\cdot}K): \textit{Confs}\to \mathit{Terms}$ is a
    surjection.
    \item \label{cor:transition} For each machine transition $K \rightarrow K'$, either
      $ \plg( \semsub K K) = \plg( \semsub {K'} K)$ (\ie, the two
      configurations represent different decompositions of the same
      term), or $ \plg( \semsub K K)$ reduces to
      $\plg( \semsub {K'} K)$ in the strategy.
    \item \label{cor:silent} There are no silent loops in the machine,
      \ie, no infinite sequences of transitions
      $K_0\rightarrow \ldots \rightarrow K_n \rightarrow \ldots$ such
      that $\plg(\semsub{K_i}K)=\plg(\semsub{K_{i+1}} K)$ for all $i$.
    \item \label{cor:reduction} For all terms $T,T'$, if $T$ reduces
      to $T'$ according to the strategy, then for each $K$ such that
      $\plg(\semsub K K)=T$ there exists a sequence of machine
      transitions $K \rightarrow \ldots \rightarrow K'$ such that
      $\plg(\semsub {K'} K)=T'$.
    \end{enumerate}
\end{theorem}
The proof of this theorem is more tedious than sophisticated. We
provide a sketch in Appendix~\ref{app:correctness}. 
Point~\ref{cor:surj} is a simple
observation that for any term $T$ the corresponding initial
configuration is decoded to $T$. 
For point~\ref{cor:transition}, a simple case analysis gives that all
transitions $\stackrel{\neq(\ref{tr:6})}{\to}$ leave the decoding of
the configurations unchanged. 
The fact that $\stackrel{(\ref{tr:6})}{\to}$ implements
$\bwnf$-contraction is technically more involved, but not surprising.

Probably the most interesting part concerns point~\ref{cor:silent},
which implies that the machine always finds a redex in a finite number
of steps.  We present the main intuitions here, leaving formal details
in the appendices. 
We start by introducing a strict
partial order on annotated terms and frames.  For all terms
$T_1,\ldots, T_7$ we set
$$T_1 < \flapp {T_2} < \frapp {T_3} < \strongWannotation{T_4} < \flstrongann {T_5} < \frstrongann {T_6} < \flam
< \strongNannotation{T_7}$$
Then we extend this order to the reversed lexicographic extension
$\rlex$ of $<$ on annotated decompositions: $D_1\rlex D_2$ iff
$D_1^R\lex D_2^R$ where $D^R$ denotes the reverse of $D$ and $\lex$ is
the standard lexicographic extension of $<$.  Since a given term may
have only finitely many corresponding annotated decompositions that
cannot grow forever, there are no silent loops.

\begin{example}\label{ex:incr}
  The following is the sequence of decodings of configurations from
  Example~\ref{ex:evalomega}. Note that this sequence is strictly
  increasing in the $\rlex$ order.
\begin{align*}
\stackthree{\strongWannotation{0}}{\frapp{0}}{\flam}
\;\stackrel{(\ref{tr:7})}{\to} &
\;\stacktwo{\strongWannotation{\tappp{0} {0}}} {\flam}
\;\stackrel{(\ref{tr:9})}{\to}
\;\stackthree {\strongWannotation{0}} {\flstrongann {0}}{\flam}
\;\stackrel{(\ref{tr:10})}{\to}\\
\;\stackthree {\strongNannotation{0}} {\flstrongann {0}}{\flam}
\;\stackrel{(\ref{tr:13})}{\to}&
\stackthree {\strongWannotation{0}}{\frstrongann {0}}{\flam}
\end{align*}
\end{example}
%

\subsection{Corollaries}

Since all the transformations used in the derivation are
meaning-preserving, we can informally state that:
For every closed term $T$ and its OCaml representation~{\tt t}, the
computation {\tt eval t [] 0} in the call-by-value normalizer of
Figure~\ref{fig:nbe-cbv} returns a {\tt sem} value iff $T$ reaches
weak normal form in the strategy.
Similarly the computation {\tt nbe t} returns a {\tt term} value {\tt t'}
iff $T$ reaches a normal form $T'$ in the
strategy, and {\tt t'} is an OCaml representation of $T'$.


\section{Conclusion and future work}
\label{sec:conclusion}

We presented the first systematic derivation of an abstract machine
KNV that implements the strong CbV strategy for normalization in the
lambda calculus. The derivation starts from the KN machine for
normal-order reduction and uses off-the-shelf tools to transform
semantic artefacts in a sequence of steps that constitute the
so-called functional correspondence, as a two-way derivation recipe.
We also presented the reduction semantics for the strong CbV strategy
that can be read off the obtained machine, and that is an example of a
hybrid strategy with three kinds of reduction contexts. As an example
application of the machine, we illustrated how it can be used for
convertibility checking, \eg, in proof assistants based on
dependent type theory.






In~\cite{DBLP:journals/scp/AccattoliG19}, the authors introduced a
time complexity criterion for an abstract machine: a machine is called
a~\textit{reasonable implementation} of a~given strategy if it can
simulate $n$ reduction steps in a number of transitions that is
polynomial  in $n$ and in the size of the initial term. It is easy
to observe that KNV is not a reasonable implementation of strong CbV
due to the size explosion problem. Consider, \eg, the following term
family $e_n$ where $c_n$ denotes the $n$th Church numeral:

$$ \omega := \tlam{x}{\tapp{x}{x}} \hspace{3cm}
e_n := \tlam{x}{\tapp{\tapp{c_n}{\omega}}{x}}$$

\noindent
Each $e_n$ reduces to its normal form in the number of steps linear in
$n$, but the size of this normal form is exponential in $n$. Since KNV
never reuses structures constructed before, it has to introduce each
of the exponentially many constructors in a separate step. Therefore,
it is not a reasonable implementation. We intend to construct a
modified version of KNV that will critically rely on sharing of
intermediate results. We conjecture that such a modification is both
necessary and sufficient to achieve a reasonable implementation of
strong CbV. We also believe that the present development is a crucial
stepping stone in this undertaking and that it offers all the
necessary tools. In particular, sharing, in more than one flavour, can
be most naturally introduced at the level of the evaluator of
Figure~\ref{fig:nbe-cbv} and the resulting abstract machine will be a
reflection of this modification through the functional correspondence.

\paragraph{Acknowledgements.}
We thank Filip Sieczkowski and the anonymous reviewers for their
helpful comments on the presentation of this work.

\bibliography{mybib-short}
\bibliographystyle{splncs04}


\newpage
\appendix

\section{Decoding of machine representations}
\label{app:decoding} 
\subsubsection*{Terms.}
The functions
$\semsub {\cdot, \cdot} W : \mathit{Wnfs} \to \mathbb N \to \setofwnf$ and 
$\semsub {\cdot, \cdot} I : \mathit{Inerts} \to \mathbb N \to \setofinert$
decode
the machine representations of weak normal forms and  the representations of inert terms.
\begin{align*}
\semsub{\abstr T E, m} W &= \lambda{\semsub{T, \cons {\lvar{m+1}}  E, m+1} T}\\
\semsub{I, m} W &= {\semsub{I, m} I}
\\
\semsub{\lvar n, m} I &= m - n\\
\semsub{\tappp I W, m} I &= \tappp{\semsub{I, m} I}{\semsub{W, m} W}.
\end{align*}

The function
$\semsub{\cdot, \cdot, \cdot} T : \textit{Terms} \to \textit{Envs} \to
\mathbb N \to \textit{Terms}$ decodes machine representations of
arbitrary terms; it uses an auxiliary function
$\semsub {\cdot, \cdot, \cdot} n : \mathbb N \to \textit{Envs} \to
\mathbb N \to \textit{Terms}$ that implements a lookup of a variable
in an environment.
\begin{align*}
\semsub{\tappp {T_1}{T_2}, E, m} T &= \tappp
{\semsub{T_1, E, m} T}
{\semsub{T_2, E, m} T}\\
\semsub{\lambda T, E, m} T&= \semsub{\abstr T E, m} W\\
\semsub{n, E, m} T&= \semsub{n, E, m} n\\
\semsub{0, \cons W E, m} n&= \semsub{W, m} W \\
\semsub{n+1, \cons W E, m} n&= \semsub{n, E, m} n
\end{align*}
\subsubsection*{Stacks.}
\begin{align*}
\semsub \nil {S} &= \nil\\
\semsub {\cons \flam {S}} {S} &= \cons \flam \semsub {S}{S} \\
\semsub {\cons {\flstrong I} {S}} {S} &= \cons {\flstrongann{\semsub{I, |S|_\flam}I}} \semsub {S}{S}\\
  \semsub {\cons{\frstrong{\nf}}{S}} {S} &= \cons {\frstrongann \nf} \semsub {S}{S} \\
\semsub {\cons{\flclos{T}{E}}{S}} {S} &= \cons{\flapp {\semsub {T, E, |S|_\flam} T}} \semsub {S}{S}\\
\semsub {\cons{\frapp{W}}{S}} {S} &= \cons {\frapp {\semsub {W, |S|_\flam} W}}\semsub {S}{S}
\end{align*}
\subsubsection*{Configurations.}
\begin{align*}
\semsub{\wconf T E {S} m} K &= \cons {\semsub{T, E, m} T}{\semsub {S} {S}} \\
\semsub{\bconf W {S}} K &= \cons {\strongWannotation{\semsub{W, m} W}}\semsub{S}{S} \\
\semsub{\sconf T m {S}} K &= \cons{\strongNannotation{T}}\semsub{S}{S} \\[1ex]
                   \plg(\cons T \nil) &= T\\
  \plg (\cons{T_1}{\cons{\flapp {T_2}}L}) &= \plg (\cons{\tapp {T_2}{ T_1}}L)\\
  \plg (\cons{T_1}{\cons{\frapp {T_2}}L}) &= \plg (\cons{\tapp {T_1} {T_2}}L)\\
    \plg (\cons{T}{\cons{\flam}L}) &= \plg (\cons{\tolam T}L)
\end{align*}

\section{Closedness invariants}
\label{app:closedness}
The following function defines the number of lambda constructors
needed to close a given term.
\begin{align*}
\openT n &= n + 1\\
\openT {\tappp {T_1} {T_2}} &= \max\{ \openT {T_1}, \openT {T_2} \}\\
\openT {\tolam T} &= \max\{ \openT T -1, 0\}
\end{align*}

Obviously a term $T$ is closed if $\openT T = 0$. We say that an
environment $E$ \emph{closes} $T$ if $\openT T \leq |E|$. We then extend
the function to (machine representations of) environments and weak
normal forms.

\begin{align*}
\openW {\abstr T E} &= \openE E \text{\;\;if $E$ closes $\tolam T$}\\
\openW I &= \open I I\\
\openI {\lvar n} &= n\\
\openI {\tappp I W} &= \max\{ \openI I, \openW W\}\\
\openE \nil &= 0\\
\openE {\cons W E} &= \max\{ \openW W, \openE E\}
\end{align*}

\begin{lemma}\label{lem:inv}The machine maintains following invariants:
  
\begin{tabular}{rcl}
In configurations $\wconf T E S m$ &\;\;& $T$ is closed by $E$ and $\openE E \leq m = |S|_\flam$\\
In configurations $\bconf W S$ &&  $\openW W \leq m =|S|_\flam$\\
In configurations $\sconf T m S$ && $\openT T \leq m = |S|_\flam$\\
In stacks $\cons{\frapp W} S$ && $\openW W \leq |S|_\flam$\\
In stacks $\cons{\flstrong I} S$ && $\openI I \leq |S|_\flam$\\
\end{tabular}
\end{lemma}

Invariants from Lemma~\ref{lem:inv} are needed in proofs of lemmas
involving reachable configurations like
Lemma~\ref{lem:contraction}. Unreachable configurations may be subject
of strange anomalies. For example, decoding of
$\semsub{ \abstr{1}{\nil}, 0} W$ does not exist because $\tolam{1}$ is
an open term and $\nil$ does not close it. A weird thing happens when
one tries to decode
$\semsub{ \tolam{\tolam 2}, \cons{\lvar{5}} {\nil}, 4}T$: here
$\tolam {\tolam 2}$ is an open term
but it is decoded to $\tolam {\tolam 1}$,
which is a closed term. In an appropriate context this may lead to a
violation of a variant of Lemma~\ref{lem:contraction} for
non-reachable configurations:




$$\tappp {\semsub{ \tolam {\tolam 2}, \cons{\lvar{5}\!} {\!\nil}, 4 } T} {\semsub{ {\lvar 2}, 4 } W} =
\tappp {(\tolam {\tolam 1})} {2}
\to_\beta \tolam 3
\neq -1
= \semsub{\tolam 2, \cons{\lvar{2}\!} {\cons{\!\lvar{5}\!} {\!\nil}}, 4} T$$


\section{Sketch of correctness proof}
\label{app:correctness}
Below we state the main lemmas needed in the proof of Theorem~\ref{thm:correctness}.

\begin{lemma}[initial correctness]\label{lem:initial} If $T$ is a
  closed term then $ \semsub{T, \nil, 0} T = T$.
\end{lemma}

\begin{lemma}[overhead identification]
  \label{lem:plug}
  If $K$ is a reachable configuration and
  $K \stackrel{\neq(\ref{tr:6})}{\to} K'$ then
  $ \plg( \semsub K K) = \plg( \semsub {K'} K)$.
\end{lemma}

\begin{lemma}[correctness of contraction]\label{lem:contraction}
  If $\bconf {\abstr T E} {\cons {\frapp W} {S}}$ is a reachable
  configuration of the machine then
  $\tappp {\semsub{\tolam T, E, m} T} {\semsub{W, m} W}
  \rightharpoonup_\beta {\semsub{T, {\cons {W\!}{\! E}}, m} T.} $\end{lemma}

\begin{lemma}[strategy simulates machine] \label{lem:strsim} If $K$ is a reachable
  configuration of the machine and $K \stackrel{(\ref{tr:6})}{\to} K'$
  then
  $ \plg( \semsub K K) \stackrel{R}{\to}_{\bwnf} \plg( \semsub {K'}
  K)$.
\end{lemma}
\begin{proof} By correct contraction implementation and decoding of stacks. \end{proof}

The following lemma formalizes the intuition from Example~\ref{ex:incr}.
\begin{lemma}\label{lem:incr}
  If $K_1\to K_2$ by any of the transitions  (\ref{tr:1})--(\ref{tr:2}), (\ref{tr:5}),
  (\ref{tr:7})--(\ref{tr:14}) then $\semsub{K_1} K \rlex \semsub{K_2} K$.
\end{lemma}
\begin{proof}
  The proof is done by inspection of the respective transitions. Below
  we summarize, for each of the involved transitions, the key arguments.
For all terms $T, T_1,T_2$ and all lists of annotated contexts $L$:
  \begin{align}
\rightcounter{tr:1}
\cons{\tapp {T_1} {T_2}}L  &\rlex \cons {T_2} {\cons{\flapp {T_1}}L} \\
\cons{\tolam T} L &\rlex \cons{\strongWannotation {\tolam T }}L\\
\rightcounter{tr:5}
\cons{\strongWannotation{T_2}} {\cons{\flapp {T_1}} L} &\rlex \cons{T_1}{\cons{\frapp{T_2}}L}\\
\rightcounter{tr:7}
 \cons{\strongWannotation{T_1}}{\cons{\frapp{T_2}}L}  &\rlex \cons{\strongWannotation{\tappp {T_1} {T_2}}}L\\
 \cons{\strongWannotation{\tolam T}}L  &\rlex \cons T {\cons{\flam}L}\\
\cons{\strongWannotation{\tapp {T_1} {T_2}}}L  &\rlex \cons{\strongWannotation{T_2}} {\cons{\flstrongann {T_1}}L} \\
\cons{\strongWannotation {n}} L &\rlex  \cons{\strongNannotation{n}}L\\
\rightcounter{tr:13}
 \cons{\strongNannotation{T_2}} {\cons{\flstrongann {T_1}}L}    &\rlex  \cons{\strongWannotation{T_1}}{\cons{\frstrongann {T_2}}L}\\ 
\cons{\strongNannotation{T}} {\cons{\flam}L} &\rlex \cons{\strongNannotation{\tolam T}}L\\
\cons{\strongNannotation{T_1}}{\cons{\frstrongann {T_2}}L} &\rlex  \cons{\strongNannotation{\tapp {T_1} {T_2}}}L
\end{align}
\end{proof}
The following lemma formally states the main argument behind the ``no silent loops'' result.
\begin{lemma}[no silent loops]\label{lem:termination}
  Every sequence of transitions not involving the
  transition~$\stackrel{(\ref{tr:6})}{\to}$ is finite.
\end{lemma}
\begin{proof}
  Consider any sequence $K_1,K_2,\ldots$ of transitions such that
  $K_i \stackrel{\neq(\ref{tr:6})}{\to} K_{i+1}$ for all $i$. By
  Lemma~\ref{lem:plug} we have
  $ \plg( \semsub {K_i} K) = \plg( \semsub {K_j} K)$ for all $i,j$.
  Since for a given term there exist only finitely many possible
  decompositions into elementary contexts and subterms, there exist
  only finitely many annotated decompositions $D$ such that
  $\plg (D) = \plg( \semsub {K_1} K)$.

  Machine transitions (\ref{tr:3})--(\ref{tr:4}) leave unchanged the
  decoding of a configuration to an annotated decomposition, in
  symbols if $K_i\stackrel{(\ref{tr:3},\ref{tr:4})}{\to} K_{i+1}$ then
  $\semsub {K_i} K=\semsub {K_{i+1}} K$. However, the number of
  consecutive transitions (\ref{tr:3})--(\ref{tr:4}) is bounded by the
  biggest de Bruijn index (which is bounded by the number of $\lambda$
  constructors) in $ \plg( \semsub {K_i} K)$. All other transitions
  strictly increase the decoding $\semsub {K_i} K$ by
  Lemma~\ref{lem:incr}.  Therefore the sequence is finite.
\end{proof}

The remaining lemmas in this section
(Lemmas~\ref{lem:determinism}--\ref{lem:machsim}) are key observations
in the proof of point~\ref{cor:reduction} of
Theorem~\ref{thm:correctness}.

\begin{lemma}[strategy determinism]\label{lem:determinism}
  Any term has at most one ${\bwnf}$-redex positioned in $R$ context.
\end{lemma}

For the sake of proof we will say that a term has a $C$-decomposition
if it has a $\bwnf$-redex positioned in a context $C$.
We will use a fact that any term has at most one $F$-decomposition
and if it has no $F$-decomposition then it is a weak normal form.
We will prove stronger lemma than Lemma~\ref{lem:determinism}:

\begin{lemma}Every term has at most one $R$-decomposition.
If it has no $R$-decomposition then it is a normal form.\end{lemma}

\begin{proof}
Proof by structural induction on term $T$.

1. If term is a variable it has no redex and is a normal form.  2. If
term is an abstraction then its body either has at most one
$R$-decomposition or is a normal form by induction hypothesis, and so
is $T$.  3. Otherwise it is an application: $T = \tappp {T_1} {T_2}$.

3.1. If its right branch $T_2$ has a $F$-decomposition then
no redex can be found in the left branch $T_1$
because $T_2$ can't be peeled of as a $\frapp \wnf$ nor $\frstrong \nf$ frame.
3.2. Otherwise right branch $T_2$ is a weak normal form.

3.2.1. If the left branch $T_1$ has a $F$-decomposition then
no redex can be found in the right branch $T_2$
because $T_1$ can't be peeled off as a $\flstrong \inert$ frame.
3.2.2. Otherwise left branch $T_1$ is a weak normal form.

3.2.2.1. If the left branch $T_1$ is an abstraction
this application raises $\bwnf$-redex.
No redex can be found in the right branch $T_2$
because it has no {$F$-decomposition}
and abstraction $T_1$ can't be peeled off as a $\flstrong \inert$ frame.
No redex can be found in the left branch $T_1$
because abstraction has no $F$-decomposition or $H$-decomposition.
3.2.2.2. Otherwise $T_1$ is an inert term.

3.2.2.2.1. If $T_2$ has a $R$-decomposition then
no redex can be found in the left branch $T_1$
because $T_2$ can't be peeled of as a $\frstrong \nf$ frame.
It's the only $R$-decomposition of $T_2$ by induction hypothesis.
3.2.2.2.2. Otherwise $T_2$ is a normal form by induction hypothesis.

3.2.2.2.2.1. If $T_1$ has a $R$-decomposition then
it's the only $R$-decomposition by induction hypothesis.
3.2.2.2.2.2. Otherwise $T_1$ is a neutral term by induction hypothesis
and so $T$ is.

\end{proof}

\begin{lemma}[machine simulates strategy]\label{lem:machsim}
  If $\plg(\semsub{K} K)$ has a ${\bwnf}$-redex positioned in $R$
  context then there exists $K'$ s.t.
  $K \to^*\stackrel{(\ref{tr:6})}{\to} K'$.
\end{lemma}

\begin{proof}The only way machine can stop is to build a normal
  form. From Lemma~\ref{lem:termination}. if term has a redex the
  machine must perform its contraction.\end{proof}


\end{document}